\documentclass[12pt,a4paper]{article}
\usepackage{times}
\usepackage{graphics}
\usepackage{graphicx}
\usepackage{float}
\restylefloat{figure}

\newcommand{\si}{\mbox{\boldmath $\sigma$}}

\newcommand{\B}{\mbox{\boldmath $B$}}
\newcommand{\mb}{\mbox{\boldmath $\mu$}}

\begin{document}
\date{ }

\begin{center}
{\Large Potential of the neutron Lloyd's mirror interferometer

for the search for new interactions}

\vskip 0.7cm

Yu. N. Pokotilovski\footnote{e-mail: pokot@nf.jinr.ru}

\vskip 0.7cm
            Joint Institute for Nuclear Research\\
              141980 Dubna, Moscow region, Russia\\
\vskip 0.7cm

{\bf Abstract\\}

\begin{minipage}{130mm}

\vskip 0.7cm
 We discuss the potential of the neutron Lloyd's mirror
interferometer in a search for new interactions at small scales.
 We consider three hypothetical interactions that may be tested using the
interferometer.
 The chameleon scalar field proposed to solve the enigma of accelerating
expansion of the Universe produces interaction between particles and matter.
 The axion-like spin-dependent coupling between neutron and nuclei or/and
electrons may cause P- and T-non-invariant interaction with matter.
 Hypothetical non-Newtonian gravitational interactions mediates additional
short-range potential between neutrons and bulk matter.
 These interactions between the neutron and the mirror of the Lloyd's type
neutron interferometer cause phase shift of neutron waves.
 We estimate the sensitivity and systematic effects of possible experiments.
\end{minipage}
\end{center}
\vskip 0.3cm

PACS: 14.80.Mz;\quad 12.20.Fv;\quad 29.90.+r;\quad 33.25.+k

\vskip 0.2cm

Keywords: Chameleon scalar fields; Axion; Non-Newtonian gravity; Neutron interferometers

\vspace{5mm}

{\bf 1. Introduction}

 It is believed that the Standard model is a low energy approximation of some
more fundamental theory.
 Most popular extensions of the Standard Model: supersymmetry and superstring
theories, predict the existence of new particles and hence new interactions.
 These new particles were not detected up to now because of their too large
mass, or because of too weak interaction with ordinary matter.
 This last case is of interest in our discussion of a search for
new hypothetical weak interactions of different nature.

 The possible existence of new interactions with macroscopic ranges and weak
coupling to matter currently attracts increasing attention.
 Significant number of experiments has been performed to search for new forces
in a wide range of distance scales.
 Here we consider possibilities of the neutron Lloyd's mirror interferometer in
searching for some of these new interactions.

 The Lloyd's mirror interferometer (see Fig. 1) well known in the light optics
has not yet been discussed in the experimental neutron optics.

 The geometric phase shift is determined by the difference of path lengths
of the reflected and non-reflected beams:
\begin{equation}
\varphi_{geom}=\varphi_{II,geom}-\varphi_{I,geom}=
k\Bigl(\sqrt{L^{2}+(b+a)^{2}}-\sqrt{L^{2}+(b-a)^{2}}\Bigr)\approx 2kab/L,
\end{equation}
where $k$ is the neutron wave vector, $L$, $b$ and $a$ are given in the Fig. 1
caption.
 The last equation is valid with relative precision better than $ab/L^{2}$.
 The geometric phase shift linearly depends on the interference coordinate $b$.
 It means that the interference pattern $I\sim sin^{2}(\pi ab/\lambda_{n}L)$
in the absence of any potentials is sinusoidal with high precision:
 $ab/L^{2}\sim 10^{-8}$ at $a\sim b\sim 10^{-2}$\,cm and $L=1$\,m.
 The period of oscillations in the interference pattern is
$\Lambda_{osc}=\lambda_{n}L/(2a)$, where $\lambda_{n}$ is the neutron
wavelength, and is $\sim 1\,\mu$m for the thermal neutrons energy range and
reasonable parameters of the interferometer.
 But for very cold neutrons in the $\mu$eV-energy range the period of the
interference oscillations approaches dozens of $\mu$m, and an interference
picture may be registered with a narrow (about $\sim 1\mu$m) slit at a detector
window or with modern high resolution position sensitive neutron detectors.

 The idea of possible application of the Lloyd's mirror interferometer for the
search for new hypothetical interaction between matter and particles consists
in measuring the neutron wave phase shift produced by a hypothetical
mirror-neutron potential.
 We here consider three actively discussed hypothetical interactions:
the cosmological scalar fields proposed to explain the accelerated expansion of
the Universe, the axion-like spin-dependent pseudoscalar nucleon-nucleon and/or
nucleon-electron interaction, and hypothetical deviation of the gravitation law
from the Newtonian one at small distances (non-Newtonian gravity).

\vspace{5mm}

{\bf 2. Chameleon scalar field}

 There is evidence of the accelerated expansion of the Universe.
 The nature of this effect is one of the most exciting problems in physics and
cosmology.
 It is not clear yet whether the explanation of the fact that gravity becomes
repulsive at large distances should be found within General Relativity or
within a new theory of gravitation.
 One possibility to explain this fact is to modify the General Relativity
Theory, and there was a number of proposals of this kind.
 Among various ideas proposed to explain this astronomical observation in a
different way, one of popular variants is a new matter component of the
Universe -- a cosmological scalar field of the quintessence type \cite{Ratra}
dominating the present day density of the Universe (for the recent reviews see
\cite{Pee,Cop}).

 Acting on cosmological distances the mass of this field should be of the order
of the Hubble constant: $\hbar H_{0}/c^{2}=10^{-33}\,eV/c^{2}$.

 The massless scalar fields appearing in string and supergravity theories
couple to matter with gravitational strength.
 Because of direct coupling to matter with a strength of gravity, the existence
of light scalar fields leads to a violation of the equivalence principle.
 In the absence of self-interaction of the scalar field, the experimental
constraints on such a field are very strict, requiring its coupling to matter
to be unnaturally small.

 The solution proposed in \cite{Kho,Bra04,Gub,Upa06,Mota06,Mota07} consists in
the introduction of the coupling of the scalar field with matter of such a form
that as a result of self-interaction and the interaction of the scalar field
with matter, the mass of the scalar field depends on the local matter
environment.

 In the proposed variant, the coupling to matter is of the order as expected
from string theory, but is very small on cosmological scales.
 In the environment of the high matter density, the mass of the field increases,
the interaction range strongly decreases, and the equivalence principle is
not violated in laboratory experiments for the search for the long-range fifth
force.
 The field is confined inside the matter screening its existence to the
external world.
 In this way the chameleon fields evade tests of the equivalence principle and
the fifth force experiments even if these fields are strongly coupled to matter.
 As a result of the screening effect the laboratory gravitational experiments are
unable to set an upper limit on the strength of the chameleon-matter coupling.

 The deviations of results of measurements of gravity forces at macroscopic
distances from calculations based on Newtonian physics can be seen in the
experiments of Galileo-, E\"otv\"os-or Cavendish-type \cite{Fisch} performed
with macro-bodies.
 At smaller distances  ($10^{-7}-10^{-2}$) cm the effect of these forces can be
observed in measurements of the Casimir force between closely placed
macro-bodies (for review see \cite{Rep}) or in the atomic force microscopy
experiments.
 Casimir force measurements may to some degree evade the screening and probe
the interactions of the chameleon field at the micrometer range despite the
presence of the screening effect \cite{Mota07,Cas,Cas1}.

 At even smaller distances such experiments are not sensitive enough, and high
precision particle scattering experiments may play their role.
 In view of absence of electric charge the experiments with neutrons are more
sensitive than with charged particles, electromagnetic effects in scattering
of neutrons by nuclei are generally known and can be accounted for with high
precision \cite{myN,NesvN}.

 As regards the chameleon interaction of elementary particles with bulk matter,
it was mentioned in \cite{BraPi} that neutron should not show a screening
effect - the chameleon-induced interaction potential of bulk matter with
neutron can be observed.
 It was also proposed in \cite{BraPi} to search for the chameleon field through
energy shift of ultracold neutrons in the vicinity of reflecting horizontal
mirror.
 From the already performed experiments on the observation of gravitational
levels of neutrons, the constraints on parameters, characterizing the force of
chameleon-matter interaction were obtained in \cite{BraPi}.

 Chameleons can also couple to photons.
 It was proposed in \cite{Ahl,Gies} to search for in a closed vacuum cavity for
the afterglow effect resulting from the chameleon-photon interaction in a
magnetic field.
 The GammeV-CHASE \cite{CHASE,CHASE1} and ADMX \cite{ADMX} experiments based on
this approach are intended to measure (constrain) the coupling of chameleon
scalar field to matter and photons.

 In the approach proposed here only the chameleon-matter interaction is measured
irrespective of the existence of the chameleon-photon interaction.
 The approach is based on the standard method of measurement the phase shift
of a neutron wave in the interaction potential.

 Testing the interaction of particles with matter at small distances may be
interesting irrespective of any particular variant of the theory.

 In one of popular variants of the chameleon scalar field theory
\cite{Kho,Bra04,Gub,Upa06,Mota06,Mota07}, the chameleon effective potential is
\begin{equation}
V_{eff}(\phi)=V(\phi)+e^{\beta\phi/M_{Pl}}\rho,
\end{equation}
where
\begin{equation}
V(\phi)=\Lambda^{4}+\frac{\Lambda^{4+n}}{\phi^{n}}.
\end{equation}
is the scalar field potential, $M_{Pl}$ is the Planck mass, $\rho$ is the local
energy density of the environment,
$\Lambda=(\hbar^{3}c^{3}\rho_{d.e.})^{1/4}$=2.4 meV is the dark energy scale,
$\rho_{d.e.}\approx 0.7\times 10^{-8}$ erg/cm$^{3}$ is the dark energy density,
and $\beta$ is the interaction parameter not predicted by the theory.

 The chameleon interaction potential of a neutron with bulk matter (mirror)
was calculated in \cite{BraPi}:
\begin{equation}
V(z)=\beta\frac{m}{M_{Pl}\lambda}\Bigl(\frac{2+n}{\sqrt{2}}\Bigr)^{2/(2+n)}
\Bigl(\frac{z}{\lambda}\Bigr)^{2/(2+n)}= \beta\cdot 0.9\cdot 10^{-21}\, eV
\Bigl(\frac{2+n}{\sqrt{2}}\Bigr)^{2/(2+n)} \Bigl(\frac{z}{\lambda}\Bigr)^{2/(2+n)}=
{V_{0}}\Bigl(\frac{z}{\lambda}\Bigr)^{2/(2+n)},
\end{equation}
\begin{equation}
V_{0}=\beta\cdot 0.9\cdot 10^{-21}\,eV \Bigl(\frac{2+n}{\sqrt{2}}\Bigr)^{2/(2+n)},
\end{equation}
where $m$ is the neutron mass and $\lambda=\hbar c/\Lambda=82\,\mu m$.


 To reduce the strong effect of Earth's gravity the mirror of the
interferometer is vertical.

 The neutron wave vector $k^{'}$ in the potential $V$ is
\begin{equation}
k^{'2}=k^{2}-\frac{2mV}{\hbar^{2}}, \qquad k^{'}=k-\frac{mV}{k\hbar^{2}}.
\end{equation}

 The phase shift due to the chameleon-mediated interaction potential of a
neutron with the mirror, depending on the distance from the mirror, is obtained
by integration along trajectories
$\varphi=\oint k^{'}ds=\varphi_{II}-\varphi_{I}$, where $\varphi_{I}$ and
$\varphi_{II}$ are the phases obtained along trajectories I and II:
\begin{eqnarray}
\varphi_{I} & = & k\sqrt{L^{2}+(b-a)^{2}}- \frac{mV_{0}\sqrt{1+((b-a)/L)^{2}}}{k\hbar^{2}\lambda^{\alpha_{n}-1}}
\int_0^L\Bigl(a+\frac{b-a}{L}x\Bigr)^{\alpha_{n}-1}dx= \nonumber\\
& & = \varphi_{I,geom}-\frac{\gamma\sqrt{1+((b-a)/L)^{2}}} {\lambda^{\alpha_{n}-1}\alpha_{n}(b-a)}
\Bigl(b^{\alpha_{n}}-a^{\alpha_{n}}\Bigr)
\end{eqnarray}
and
\begin{eqnarray}
\varphi_{II} & = & k\sqrt{L^{2}+(b+a)^{2}}-
\frac{mV_{0}\sqrt{1+((b+a)/L)^{2}}}{k\hbar^{2}\lambda^{\alpha_{n}-1}}
\Biggl[\int_0^l\Bigl(a-\frac{b+a}{L}x\Bigr)^{\alpha_{n}-1}dx
+\int_l^L\Bigl(\frac{b+a}{L}x-a\Bigr)^{\alpha_{n}-1}dx\Biggr]= \nonumber\\
& & = \varphi_{II,geom}+\frac{\gamma\sqrt{1+((b+a)/L)^{2}}} {\lambda^{\alpha_{n}-1}\alpha_{n}(b+a)}
\Bigl(b^{\alpha_{n}}+a^{\alpha_{n}}\Bigr).
\end{eqnarray}

 Here $l=(aL)/(a+b)$ is the x-coordinate of the beam II reflection point from
the mirror, $\gamma=(mV_{0}L)/(k\hbar^{2})$, and $\alpha_{n}=(4+n)/(2+n)$.

 The phase shift from the chameleon neutron-mirror potential
\begin{eqnarray}
\varphi_{cham} & = & \varphi_{II,cham}-\varphi_{I,cham}= \nonumber\\
& & = \frac{\gamma}{\lambda^{\alpha_{n}-1}\alpha_{n}}
\Bigl[\frac{b^{\alpha_{n}}-a^{\alpha_{n}}}{b-a}\sqrt{1+((b-a)/L)^{2}}-
\frac{b^{\alpha_{n}}+a^{\alpha_{n}}}{b+a}\sqrt{1+((b+a)/L)^{2}}\Bigr]
\approx \nonumber\\
& & \approx \frac{\gamma}{\lambda^{\alpha_{n}-1} \alpha_{n}}2ab
\frac{b^{\alpha_{n}-1}-a^{\alpha_{n}-1}}{b^{2}-a^{2}}.
\end{eqnarray}

 For a non-strictly vertical mirror, the component of the Earth's gravity
normal to the surface of the mirror produces the potential
$V_{gr}=\kappa mgz$, where $g$ is the acceleration of gravity, and
the coefficient $\kappa$ depends on the angle $\theta$ between
gravity vector and the mirror plane.
 At $\theta=10^{"}$, we have $\kappa\approx 5\times 10^{-5}$.
 This linear potential leads to additional phase shift
\begin{equation}
\varphi_{gr}=\varphi_{II,gr}-\varphi_{I,gr}=
\frac{\kappa gm^{2}}{2k\hbar^{2}(a+b)}\Bigl[(b+a)^{2}\sqrt{L^{2}+(b-a)^{2}}-
(b^{2}+a^{2})\sqrt{L^{2}+(b+a)^{2}}\Bigr]
\approx \frac{\kappa gm^{2}}{k\hbar^{2}}\frac{abL}{a+b},
\end{equation}
calculated in analogy with Eqs.(6)-(9).

 The Coriolis phase shift due to Earth's rotation \cite{Cori} is
\begin{equation}
\varphi_{Cor}=\frac{2m}{\hbar}(\Omega\cdot\bf A),
\end{equation}
where $\Omega$ is the vector of angular rotation of Earth and $\bf A$ is the
vector of the area enclosed by the interferometer beams.

 As $A=(abL)/(a+b)$, for the location of the Institute Laue-Langevin (where a
good very cold neutron source has been constructed \cite{Ste}) we have
$\varphi_{Cor}=0.16(abL)/((a+b))$\,rad (a,b,L in cm).
 As expected it is similar to the gravitational phase shift in its
dependence both of the slit and the interference coordinates.

 We must also calculate the phase shift of the neutron wave along beam II at
the point of reflection.
 Neglecting the imaginary part of the potential of the mirror, the amplitude of
the reflected wave is $r=e^{-i\varphi_{refl}}$, with the phase
\begin{equation}
\varphi_{refl}=2\arccos(k_{\bot}/k_{b})\approx\pi-\delta\varphi_{refl},
\end{equation}
where
\begin{equation}
\delta\varphi_{refl}=2k_{\bot}/k_{b}\approx\pi-2\frac{k}{k_{b}}\frac{a+b}{L},
\end{equation}
$k_{\bot}$ is the neutron wave vector component normal to the mirror's surface,
and $k_{b}$ is the boundary wave vector of the mirror.
 This phase shift linearly depends on $b$ similarly to the geometric phase
shift $\varphi_{geom}$.

 The reflected and non-reflected beams follow slightly different paths in the
interferometer.
 Therefore in the vertical arrangement of the reflecting mirror they spent
different times in Earth's gravitational field with $\Delta t=2ab/(Lv)$, where
$v$ is the neutron velocity.
 The difference in vertical shifts of the reflected and non-reflected beams is
$\Delta h=2gab/v^{2}$, and the phase shift due to this difference
\begin{equation}
\Delta\varphi_{vert}=kg^{2}abL/v^{4}.
\end{equation}
 With our parameters of the interferometer this value is of the order
$\sim 10^{-4}$.

 The total measured phase shift is
\begin{equation}
\varphi=\varphi_{geom}+\varphi_{cham}+\varphi_{gr}+\varphi_{Cor}+\varphi_{refl}.
\end{equation}

 The gravitational phase shift can be suppressed by installing the mirror
vertically with the highest possible precision.
 On the other hand the gravitational phase shift may be used for calibration of the
the interferometer by rotation around horizontal axis.
 The phase shifts due to Earth rotation $\varphi_{Cor}$ and reflection
$\varphi_{refl}$ may be calculated and taken into account in analysis of the
interference curve.

 Figure 2 shows the calculated phase shift $\varphi_{cham}$ for an idealized
Lloyd's mirror interferometer (strictly monochromatic neutrons, the width of the
slit is zero, the detector resolution is perfect) with parameters:
the neutron wave length $\lambda_{n}$=100 \AA, (the neutron velocity 40 m/s),
$L$=1 m, $a=100\,\mu$m, $\beta=10^{7}$, at $n$=1 and $n$=6.
 Shown also are the gravitational phase shift $\varphi_{gr}$ at
$c=5\times 10^{-5}$ (deviation of the mirror from verticality is 10''), and
$\delta\varphi_{refl}$ = $\pi-$phase shift of the ray II at reflection
($k_{b}=10^{6}$\,cm$^{-1}$).

 It is essential that the sought phase shift due to hypothetical chameleon
potential depends on the interference coordinate nonlinearly.
 Effect of the hypothetical interaction has to be inferred from analysis of the
interference pattern after subtracting off the effects of Earth gravity,
Coriolis and reflection.

 Figure 3 demonstrates the calculated interference pattern for the same
parameters of the interferometer as in Fig. 2 for two cases: (1) with the
geometrical phase shift $\varphi_{geom}$, the gravitational phase shift
$\varphi_{gr}$ at $\kappa=5\times 10^{-5}$, and the phase shift of the ray II
at reflection $\delta\varphi_{refl}=\pi - \varphi_{refl}$
($k_{b}=10^{6}$\,cm$^{-1}$) taken into account; (2) the same plus the phase
shift due to the chameleon field with matter interaction parameters
$\beta=10^{7}$,\, $n=1$.

 After subtracting all the phase shifts except purely geometrical the
interference pattern should be strongly sinusoidal with the period of
oscillations determined by the geometric phase shift:
$\Lambda_{osc}=\lambda_{n}L/(2a)$.
 The number of oscillations in an interference pattern with the coordinate less
than $b$ is $n_{osc}=2ab/(\lambda_{n}L)$.

 It follows from these calculations that the effect of the chameleon
interaction of a neutron with matter may be tested in the range of strong
coupling with the parameter of interaction down to $\beta\sim 10^{7}$ or lower.

 Existing constraints on the parameters $\beta$ and $n$ may be found in Fig. 1
of Ref. \cite{BraPi}.
 For example the allowed range of parameters for the strong coupling regime
$\beta\gg 1$ are: $50<\beta<5\times 10^{10}$ for n=1,
$10<\beta<2\times 10^{10}$ for n=2, and $\beta<10^{10}$ for $n>2$.
 It is seen that the Lloyd's mirror interferometer may be able to constrain
the chameleon field in the large coupling area of the theory parameters.

\vspace{0.3cm}

{\bf 3. Axion-like spin-dependent interaction.}

 There are general theoretical indications of the existence of interactions
coupling mass to particle spin \cite {Lei,Moh,Hill,Fay,Dob}.
 Experimental search for these forces is promising way to discover new physics.

 On the other hand, a number of concrete proposals were published of new light,
scalar or pseudoscalar, vector or pseudovector weakly interacting bosons.
 The masses and the coupling of these new hypothetical particles to nucleons,
leptons, and photons are not predicted by the proposed models.

 The popular solution of the strong CP problem is the existence of a
light pseudoscalar boson - the axion \cite{ax}.
 The axion coupling to fermions has general view $g_{aff}=C_{f}m_{f}/f_{a}$,
where $C_{f}$ is the model dependent factor.
 Here $f_{a}$ is the scale of Peccei-Quinn symmetry breaking which is not
predicted so that the axion may have a priori mass in a very large range: ($10^{-12}<m_{a}<10^{6}$) eV.
 The main part of this mass range from both -- low and high mass boundaries --
was excluded in result of numerous experiments and constraints from astrophysical considerations \cite{PDG}.
 Astrophysical bounds are based on some assumptions concerning the axion and
photon fluxes produced in stellar plasma.

 More recent constraints limit the axion mass to ($10^{-5} <m_{a}< 10^{-3}$) eV
with respectively very small coupling constants to quarks and photon \cite{PDG}.

 The axion is one of the best candidates for the cold dark matter of the
Universe \cite{dark,Pos}.

 Axions can mediate a P- and T-reversal violating monopole-dipole interaction
potential between spin and matter (polarized and unpolarized nucleons or
electrons) \cite{Mood}:
\begin{equation}
V_{mon-dip}({\bf r})=g_{s}g_{p}\frac{\hbar^{2}\si\cdot {\bf n}} {8\pi m_{n}}\Bigl(\frac{1}{\lambda
r}+\frac{1}{r^{2}}\Bigr)e^{-r/\lambda},
\end{equation}
where $g_{s}$ and $g_{p}$ are dimensionless coupling constants of the scalar and pseudoscalar vertices
(unpolarized and polarized particles), $m_{n}$ the nucleon mass at the polarized vertex, the nucleon spin ${\bf
s}=\hbar\si/2$, $r$ is the distance between the nucleons, $\lambda=\hbar/(m_{a}c)$ is the range of the force,
$m_{a}$ - the axion mass, and $\bf n={\bf r}$/r is the unitary vector.

 Several laboratory searches (mostly by the torsion pendulum method) provided
constraints on the product of the scalar and pseudoscalar couplings at
macroscopic distances $\lambda >10^{-2}$\,cm (see reviews
\cite{Adel,Adel1,Jaec,Raf1}).

 There are also experiments on the search for the monopole-dipole interactions
in which the polarized probe is an elementary particle: neutron \cite{Baes,Ser,myUCN,Ser1}, or atoms and nuclei
\cite{myHe,Fu}, correction in \cite {Pet}.

 For the monopole-monopole interaction due to exchange of the pseudo-scalar
boson \cite{Mood}
\begin{equation}
V_{mon-mon}(r)=\frac{g_{s}^{2}}{4\pi}\frac{\hbar c}{r}e^{-r/\lambda}
\end{equation}
the limit on the scalar coupling constant $g_{s}$ can be inferred from the
experimental search for the ''fifth force" in the form of the Yukawa-type
gravity potential $U_{5}(r)=\alpha_{5}GMme^{-r/\lambda}/r$:
\begin{equation}
g_{s}^{2}=\frac{4\pi Gm_{n}^{2}\alpha_{5}}{\hbar c}\approx 10^{-37}\alpha_{5},
\end{equation}
where $\alpha_{5}$ is the "fifth force" Yukawa-type interaction constant.

 It follows from the experimental tests of gravitation at small distances
(see reviews in \cite{Adel,Adel1}) that $g_{s}^{2}$ is limited by the value
$10^{-40}-10^{-38}$ in the interaction range 1 cm $>\lambda >10^{-4}$ cm.
 The sensitivity of these experiments falls with decreasing the interaction
range below $\sim$0.1 cm.

 The pseudoscalar coupling constant is restricted to $g_{p}<10^{-9}$ from
astrophysical considerations \cite{Raf,Raf1}.

 It is seen that the constraints obtained and expected from further laboratory
searches are weak compared to the limit on the product $g_{s}g_{p}<10^{-28}$
inferred from the above mentioned separate constraints on $g_{s}$ and $g_{p}$.
 Although laboratory experiments may not lead to bounds that are strongest
numerically, measurements made in terrestrial laboratories produce the most
reliable results.
 The direct experimental constraints on the monopole-dipole interaction may be
useful for limiting more general class of low-mass bosons irrespective of any
particular theoretical model.
 In what follows the constraint on product $g_{s}g_{p}$ may be used for the
limits on the coupling constant of this more general interaction.

 It follows from Eq. (16) that the potential between the layer of substance and
the nucleon separated by the distance $x$ from the surface is:
\begin{equation}
V_{mon-dip}(x)=\pm g_{s}g_{p}\frac{\hbar^{2}N\lambda}{4m_{n}} (e^{-x/\lambda}-e^{-(x+d)/\lambda}) =\pm
V_{0}e^{-x/\lambda}\qquad (d\gg\lambda),
\end{equation}
where $V_{0}=g_{s}g_{p}\hbar^{2}N\lambda/(4m_{n})$, $N$ is the nucleon density in the layer, $d$ is the layer's
thickness.
 The "+" and "-" depends on the nucleon spin projection on x-axis (the surface
normal).

 Phase shifts of beams I and II due to interaction of Eq. (20) are calculated
similarly to Eqs. (7) and (8):
\begin{eqnarray}
\varphi_{I} & = & k\sqrt{L^{2}+(b-a)^{2}}+ \frac{mV_{0}}{k\hbar^{2}L}\sqrt{L^{2}+(b-a)^{2}}
\int\limits_{0}^{L}exp\Bigl[-\Bigl(a+\frac{b-a}{L}x\Bigr)/\lambda\Bigr]dx= \nonumber\\
& & =\varphi_{geom.I}+ \frac{mV_{0}}{k\hbar^{2}}\sqrt{L^{2}+(b-a)^{2}} \frac{\lambda}{b-a}
(e^{-a/\lambda}-e^{-b/\lambda})=\varphi_{geom.I}+\varphi_{pot.I}
\end{eqnarray}
and
\begin{eqnarray}
\varphi_{II}=k\sqrt{L^{2}+(b+a)^{2}}+ \frac{mV_{0}}{k\hbar^{2}L}\sqrt{L^{2}+(b+a)^{2}}
\Biggl[\int\limits_{0}^{l}exp\Bigl[-\Bigl(a-\frac{b+a}{L}x\Bigr)/\lambda\Bigr]dx+
\nonumber\\
\int\limits_{l}^{L}exp\Bigl[-\Bigl(\frac{b+a}{L}x-a\Bigr)/\lambda\Bigr]dx\Biggr]= \varphi_{geom.II}+
\frac{mV_{0}}{k\hbar^{2}}\sqrt{L^{2}+(b+a)^{2}}\frac{\lambda}{b+a}
(2-e^{-a/\lambda}-e^{-b/\lambda})= \nonumber\\
=\varphi_{geom.II}+\varphi_{pot.II}.
\end{eqnarray}

 For the spin-dependent potential of Eq.(19) the signs of potential $V_{0}$
and, respectively, the phase shifts $\varphi_{pot}$ are opposite for two spin
orientations in respect to the mirror surface normal.

 The difference in these phase shifts, measured in the experiment
\begin{equation}
(\varphi_{I}^{+}-\varphi_{II}^{+})-(\varphi_{I}^{-}-\varphi_{II}^{-})=
2(\varphi_{I}-\varphi_{II})=\delta\varphi.
\end{equation}
 The geometric, gravitational phase shifts and phase shift of the beam II at
reflection calculated earlier do not depend on spin.

 The phase shift due to axion interaction is
\begin{equation}
\varphi_{ax}=2\gamma\lambda\Bigl[a(1-e^{-b/\lambda})-b(1-e^{-a/\lambda})\Bigr]/ (b^{2}-a^{2}),
\end{equation}
where $\gamma=g_{s}g_{p}N\lambda L/(4k)$.
 At $b=a$,\, and\, $\lambda/a\ll 1,\,\,\,\varphi_{ax}\rightarrow
\gamma\lambda/a$.

 Figure 4 shows the neutron wave phase shift $\varphi_{ax}$ for different
interaction range $\lambda$.
 Lloyd's mirror interferometer has the following parameters: neutron wave
length 100 \AA \,(neutron velocity 40 m/s), $L$=1 m, $a=100\, \mu$m, and the
interaction strength $g_{s}g_{p}=10^{-18}$.

 The possible sensitivity seen from this figure shows that constraints on the
monopole-dipole interaction which can be obtained with the method of the
neutron Lloyd's mirror interferometry is competing with best constraints
obtained by other methods (see Ref. \cite{Raf1}).

 Figure 5 shows the calculated interference pattern due to an axion-like
spin-dependent interaction.
 The gradient of the external magnetic field $\nabla(\mb\B)$ normal to the
mirror plane ($\mu_{n}$ is the neutron magnetic moment) may produce the phase
hift effect on polarized neutrons, similar to the effect of gravitational force
$F_{gr}=mg$ (see Eq. (10)).
 Simple calculation gives that magnetic field gradient 0.01 Oe/cm is
equivalent to $\sim 5\times 10^{-5}$ of the Earth gravitation.

 A significant increase in sensitivity may be achieved in the range of small
$\lambda\,\,(\lambda/a\ll 1)$ if the geometry shown in Fig. 1(2) is used, where
the slit is located in close vicinity to the surface of additional (upper),
non-reflecting mirror.
 The axion-like potential is produced in this case by both mirrors, but with
opposite signs in accordance to Eq. (20).

 To avoid multiple reflections, the boundary wave vector of the reflecting
mirror must satisfy the condition $k_{b}\le 2ka/L$, or the neutron beam
incident on the slit should be collimated so the the first half of the
reflecting mirror is not illuminated by the neutrons.
 In this geometry the phase shift due to axion-like monopole-dipole interaction
of the neutron with both mirrors is
\begin{equation}
\varphi_{ax}=2\gamma\lambda a(e^{(b-a)/\lambda}-e^{-a/\lambda}+ e^{-b/\lambda}-1)/(b^{2}-a^{2}).
\end{equation}
 In this case
$\varphi_{ax}\rightarrow\gamma
(1-e^{-a/\lambda})\rightarrow\gamma$\,\, as $b\rightarrow a$,\,\,\,
for $\lambda/a\ll 1$ compared to $\gamma\lambda/a$ for the case of
one mirror (Eq. (23).

 The gain in sensitivity at $\lambda/a\ll 1$ compared to the case of Fig. 4
is illustrated at Fig. 6.

\vspace{0.3cm}

 {\bf 4. Non-Newtonian gravity}

 New short-distance spin-independent forces are frequently predicted in
theories expanding the Standard Model.
 These interactions can violate the Equivalence Principle if they depend on the
composition of bodies, or the sort of particles.

 Precision experiments to search for deviations from Newton's inverse square
law and of violation the Weak Equivalence Principle have been performed in a number laboratories (reviews may be
found in \cite{Fisch,Adel,Adel1,Jaec,Raf1}.

 The pioneering ideas of the multi-dimensional models first formulated in
the first half of the XX-th century (G. Nordstr\" om, T. Kaluza and O. Klein)
received renewed interest in \cite{Akam,Rub,Vis}.
 The development of supergravity and superstring theories required for their
consistency extra-dimensions.
 A more recent promising development contained in
\cite{Ant1,Ark1,Ant2,Ark2,RanSun} proposed mechanisms in which the Standard
Model fields are located on the 4-dimensional brane while gravity propagates to
the (4+n)-bulk with a larger number of dimensions.
 As a result the gravitational law may be different from the Newtonian one.

 The frequently used parametrization of new spin-independent
hypothetical short-range interaction potential has the Yukawa-type form
\begin{equation}
U(r)=\frac{\alpha GMm}{r}e^{-r/\lambda},
\end{equation}
 where $G$ is the Newtonian gravitational constant, $M$ and $m$
are the masses of gravitating bodies, $\alpha$ is the dimensionless parameter
characterizing the strength of the new force relative to gravity, and
$\lambda=\hbar/(m_{0}c)$ is the Compton wave length of the particle with the
mass $m_{0}$.
 The mass $m_{0}$ can be the mass of the new scalar field responsible for the
short-range interaction.
 In this case $\alpha\sim g_{s}^{2}$ - the product of the scalar coupling
constants.
 Or the mass $m_{0}$ can be the mass of the lightest Kaluza-Klein state (which
is the leading order mode) when the short-range interaction comes from the
extra-dimensional expansion of the Standard Model.

 The strength $\alpha$ is constrained to be below unity for
$\lambda\geq 100\, \mu$m, \cite{Raf1,Kap}, but for shorter distances the
measurements are not as sensitive being complicated by the Casimir and
electrostatic forces \cite{Rep}.
 The sensitivity reached in the experiments aiming to test spin-independent
interactions between elementary particles and matter are orders of magnitude
less sensitive: at $\lambda=100\, \mu$m it is at the level $\alpha > 10^{11}$
\cite{Nes} with loss of sensitivity at lower distances.

 The potential following from the interaction of Eq. (25) between the layer of
substance and a neutron separated by the distance $x$ from the surface is:
\begin{equation}
V_{Yuk}(x)=2\pi\alpha m_{n}^{2}NG\lambda^{2}e^{-x/\lambda}=V_{0}e^{-x/\lambda},
\end{equation}
where $N\approx\rho/m_{n}$ is the nucleon density in the layer, $\rho$ is
density of the mirror, and $V_{0}=2\pi\alpha m_{n}^{2}NG\lambda^{2}$

 The potential of Eq. (26) has the same coordinate dependence as the axion-like
interaction potential of Eq. (19), therefore the expressions for the phase
shifts are similar to Eq. (23) with
$\gamma=2\pi\alpha\rho\lambda^{2}m_{n}^{2}L/(k\hbar^{2})$.

 Fig. 7 shows the phase shifts due to the non-Newtonian interaction of Eq. (26)
at the same parameters of the interferometer of Fig. 2, and
$\rho=10$ g/cm$^{3}$.

 For the "inverted" Lloyd's mirror geometry when the reflecting mirror has
much lower density so that its gravitational effect is insignificant compared
to the effect of the upper mirror the phase shift is
\begin{equation}
\varphi_{Yuk}=2\frac{\gamma\lambda}{b^{2}-a^{2}}
\Bigl[a(e^{-a/\lambda}-e^{(b-a)/\lambda})+b(1-e^{-a/\lambda})\Bigr].
\end{equation}
 At $b\rightarrow a$, and $\lambda/a\rightarrow 0$,\,\,
$\varphi_{Yuk}\rightarrow\gamma$ with significant gain in sensitivity for
$\lambda/a\ll 1$ compared to the geometry of Fig. 1(1) and Eq. (23).

 To avoid multiple reflections in this case,  the geometry of Fig. 1(3) may be
applied, in which the reflecting mirror has only half length compared to
Fig. 1(1).
 The gain ib sensitivity is illustrated in Fig. 8.

\vspace{5mm}

{\bf 5. Feasibility}

 As mentioned above, the interference may be measured step by step shifting a
narrow slit with the width $\delta b\ll\Lambda_{osc}\sim 1-5 \mu$m.
 A better option is to use the coordinate detector measuring in this way all
the interference picture simultaneously.
 The current spatial resolution of position-sensitive slow neutron detectors is
at the level of 5 $\mu$m with electronic registration \cite{eldet} and about
1 $\mu$m with the plastic nuclear track detection technique \cite{emdet}.
 With a $1 \mu$m thick $^{10}B$ neutron converter the efficiency of
registration of the 100 \AA--wave-length neutrons may approach 100 \%.

 From Ref. \cite{Ste} where the neutron phase density at the PF-2 very cold (VCN)
channel at the Institute Laue-Langevin was measured to be 0.25
cm$^{-3}$ (m/s)$^{-3}$ at v=50 m/s it is possible to estimate the VCN
flux density as $\phi_{VCN}=1.66\times 10^{5}$
cm$^{-2}$s$^{-1}$(m/s)$^{-1} \approx 1\times 10^{5}$
cm$^{-2}$s$^{-1}$\AA$^{-1}$ (at the boundary velocity of the neutron
guide 6.5 m/s).
 On the other hand Ref. \cite{Dre} gives the larger value
$\phi_{VCN}=4\times 10^{5}$ cm$^{-2}$s$^{-1}$(m/s)$^{-1}$ for the same channel.

 Using the Zernike theorem it is possible to calculate the width $d_{sl}$ of
the slit necessary to satisfy good coherence within the coherence aperture
$\omega$, i.e. the maximum angle between diverging interfering beams:
$x=\pi\omega d_{sl}/\lambda_{n}\leq 1$, if the slit is irradiated with an
incoherent neutron flux.
 As $\omega=2b_{max}/L$, the slit width
$d_{sl}\leq L\lambda_{n}/(2\pi b_{max})\approx 2\,\mu$m at $b_{max}=1$ mm
(20 orders of interference at the period of interference
$\Lambda_{osc}=\lambda_{n}L/(2a)=50 \mu$m at $a=100 \mu$m).

 At the monochromaticity 5 \AA \,(the coherence length
$l_{coh}=20\,\lambda_{n}$), the slit width $d_{sl}=2 \mu$m, the length of the
slit 3 cm, the divergence of the incident beam determined by the the VCN guide
boundary velocity of 6.5 m/s: $\Omega=6.5/100=0.065$, the interference aperture
$\omega=2b/L=0.2/100=2\times 10^{-3}$, and hence $\omega/\Omega=0.03$, we have
the count rate to all interference curve with a width of 1 mm
(20 orders of interference) is given by
$10^{5}\times 5\times 2\cdot 10^{-4}\times 3\times 0.03\sim 10$ s$^{-1}$.
 In one day measurement the number of events in one period of interference
is $\sim 4\times 10^{4}$.
 It is enough to observe the phase shift of $\sim 0.1$ corresponding
to the effect at $\beta=10^{7}$.

 Distinctive feature of the Lloyd's mirror interferometer is the possibility to
register all the interference pattern simultaneously along z-coordinate starting
from z=0 (Fig. 1).
 The measured  interference pattern is then analyzed as regards the presence of
the sought for effects, after the corrections taking the known
gravity, Coriolis, and reflection phase shifts into account.

 The LLL-type interferometers \cite{BH,RTB} may be used in principle to search
for new hypothetical interactions placing a piece of matter in the vicinity of
interfering beams.
 But the geometry of these interferometers does not permit probing hypothetical
short-range interactions: the axion-like or non-Newtonian gravity in this way.

 We may estimate sensitivity of the LLL-type interferometer to the chameleon
potential, which is actually not short-range.
 The phase shift in this case is
\begin{equation}
\varphi_{LLL}=\frac{2\gamma\sqrt{1+(2a/L)^{2}}}{\lambda^{\alpha_{n}-1}}
(2a)^{\alpha_{n}-1},
\end{equation}
where $a$ is the half distance between the beams of the LLL-interferometer.
 The sensitivity of the Lloyd's mirror and the LLL-interferometers is
determined by the factor $La^{\alpha_{n}-1}/k$, which is much in favor of the
Lloyd's mirror interferometer.

 In the interferometers of the LLL-type \cite{BH,RTB} an interference pattern
is obtained point by point by rotation of a phase flag introduced into the
beams.
 In the case of the VCN three grating interferometers (for example
\cite{Io,Eder,Zo}) the phase shift between the beams is realized by the
same method or by shifting position of the grating.

 The neutron Lloyd's mirror experiments may be performed with monochromatic
very cold neutrons, or in the time-of-flight mode using large wave length
range, for example 80-120 \AA.
 The pseudo-random modulation \cite{corr1,corr2} is used in the correlation
time-of-flight spectrometry.
 It was realized in the very low neutron energy range \cite{corr3}.
 In this case a two-dimensional interference coordinate -- time-of-flight
registration gives significant statistical gain.

 As in the VCN interferometers based on three gratings, in the Lloyd's mirror
neutron interferometer the space between the beams is small (parts of mm).
 Therefore it hardly can be used in experiments where some devices are
introduced in the beams, or between the beams (for example to investigate
non-local quantum-mechanical effects).
 But it may be applicable to search for short-range interactions when they are
produced by a reflecting mirror.

\vspace{5mm}

{\bf 6. Acknowledgement}

 The author is indepted to an anonymous referee for the comments on the first
version of the manuscript and suggestions.

\newpage

\begin{figure}
\begin{center}
\resizebox{13cm}{13cm}{\includegraphics[width=\columnwidth]{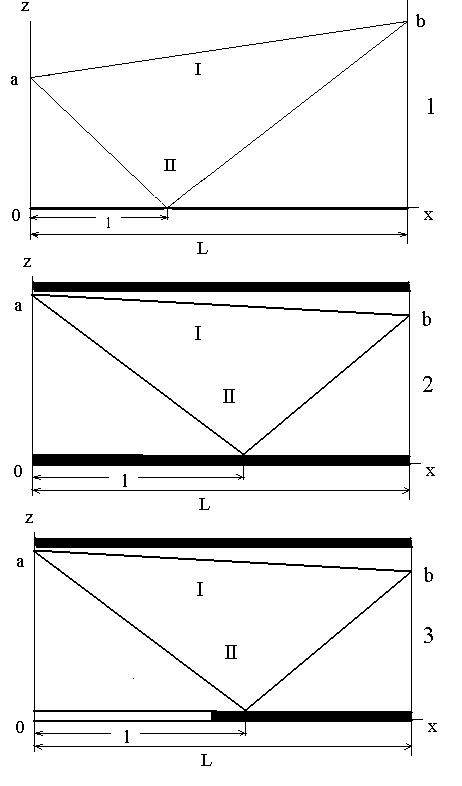}}
\end{center}
\caption{Three possible configurations of the neutron Lloyd's mirror
interferometer.
1 -- the standard Lloyd's mirror geometry,
2 -- interferometer with two mirrors, only the bottom one is reflecting,
3 -- the length of the reflecting mirror is decreased twice to avoid
multiple reflections.
 The height of the slit above the reflecting plane is $a$, $L$ is the distance
from the slit to the detector surface, $b$ is the distance of the detector
coordinate from the reflecting plane.}
\end{figure}

\newpage

\begin{figure}
\begin{center}
\resizebox{13cm}{13cm}{\includegraphics[width=\columnwidth]{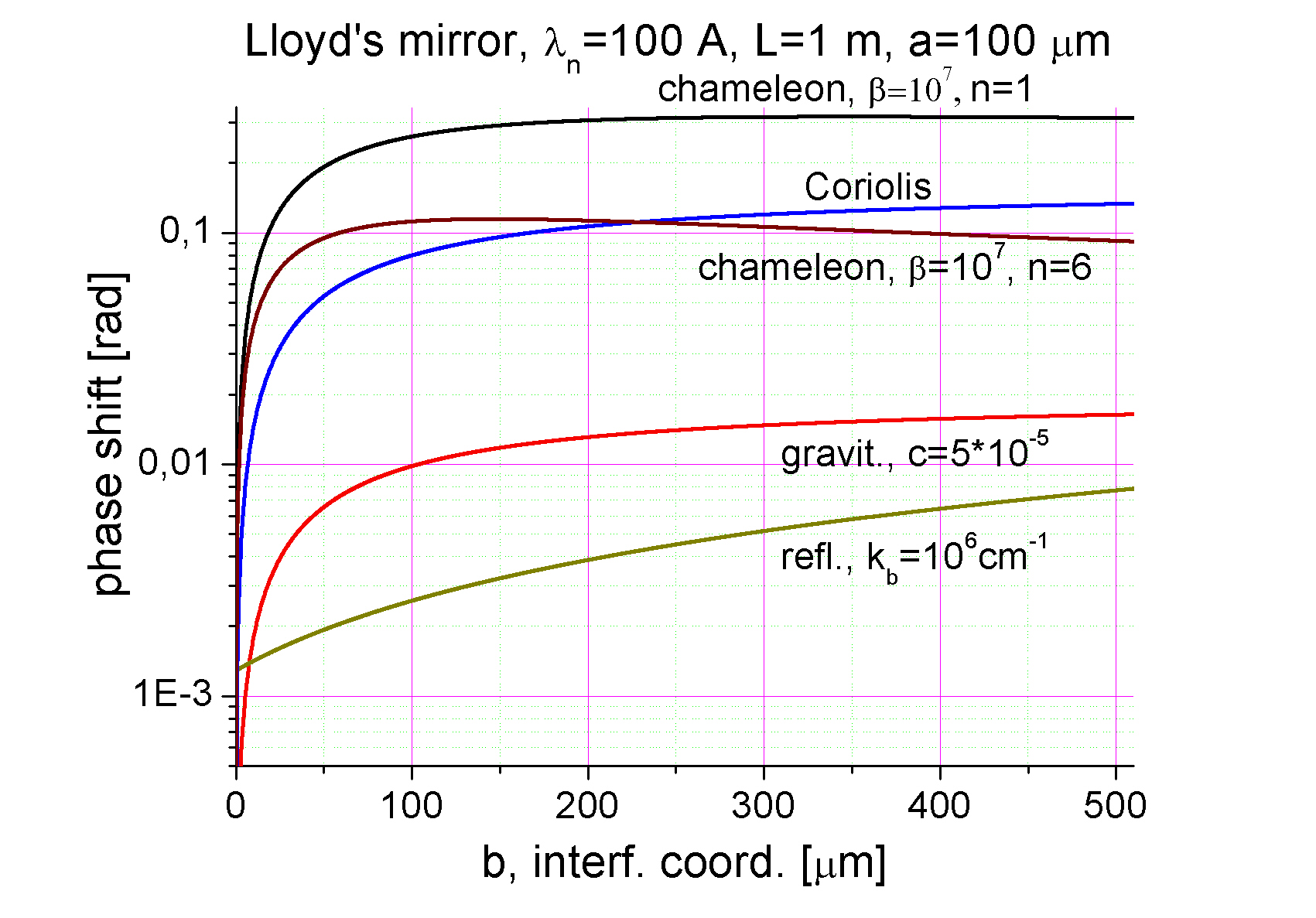}}
\end{center}
\caption{The neutron wave phase shifts $\varphi_{cham}$ in the Lloyd's mirror
interferometer with parameters:
the neutron wave length 100\,\AA, $L$=1 m, $a=100\,\mu m$, the interaction
parameters of the chameleon field with matter $\beta=10^{7}$,\, $n=1$ and $n=6$.
Also shown: the gravitational phase shift $\varphi_{gr}$ at
$\kappa=5\times 10^{-5}$; the Coriolis phase shift $\varphi_{Cor}$ ,
and the effect of reflection
$\delta\varphi_{refl}=\pi - \varphi_{refl}$ at $k_{b}=10^{6}$\,cm$^{-1}$.
 The period of oscillations in the interference pattern is
$\Lambda_{osc}=\lambda_{n}L/(2a)=50 \mu$m.}
\end{figure}

\newpage

\begin{figure}
\begin{center}
\resizebox{13cm}{13cm}{\includegraphics[width=\columnwidth]{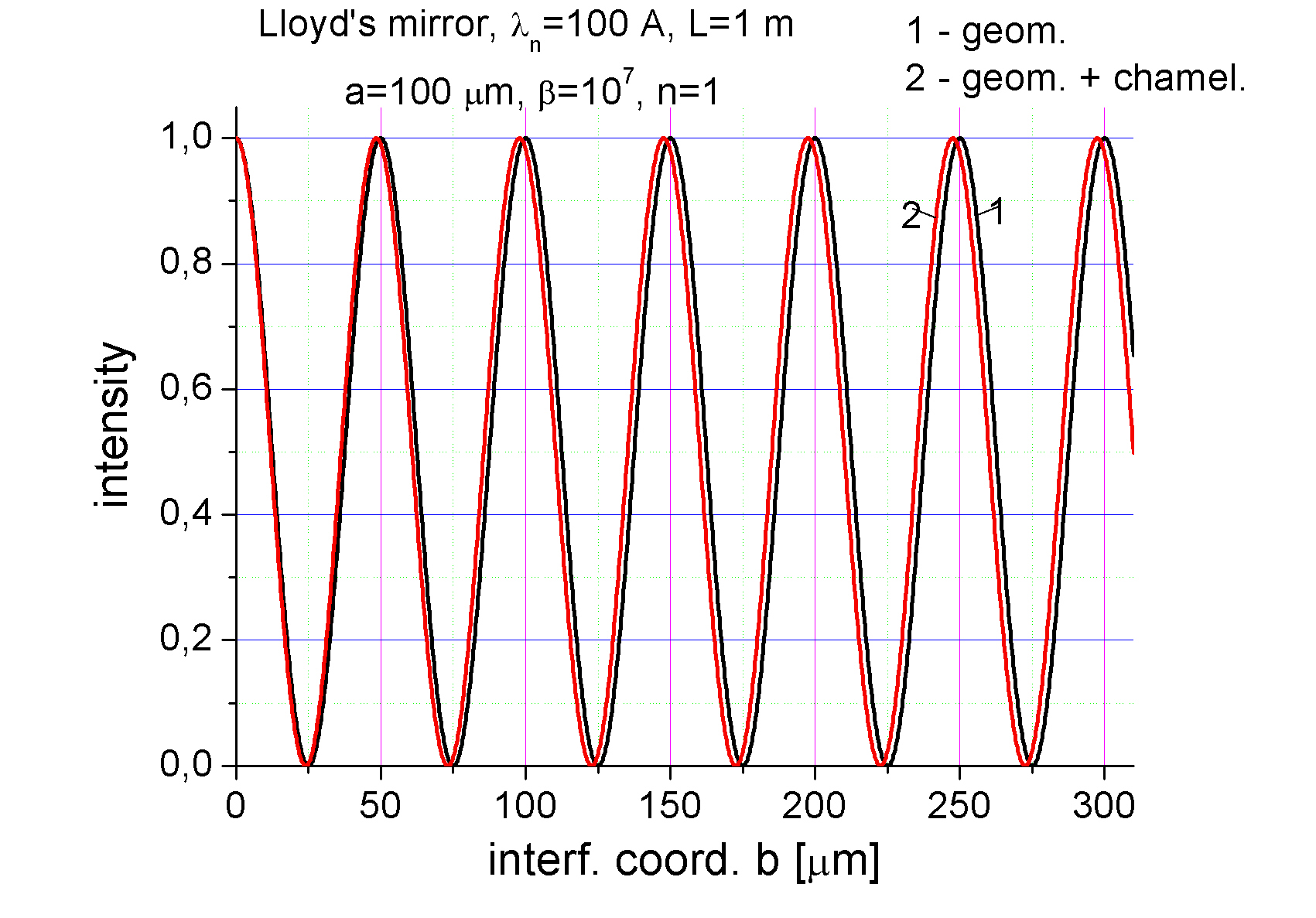}}
\end{center}
\caption{The calculated interference pattern for the neutron-mirror interaction
via the chameleon field (the parameters of the interferometer are the same as
in Fig. 2). 1 -- purely geometrical phase shift $\varphi_{geom}$,
2 -- the geometrical phase shift plus the phase shift due to the chameleon field
with matter interaction parameters $\beta=10^{7}$,\,  $n=1$.}
\end{figure}

\newpage

\begin{figure}
\begin{center}
\resizebox{13cm}{13cm}{\includegraphics[width=\columnwidth]{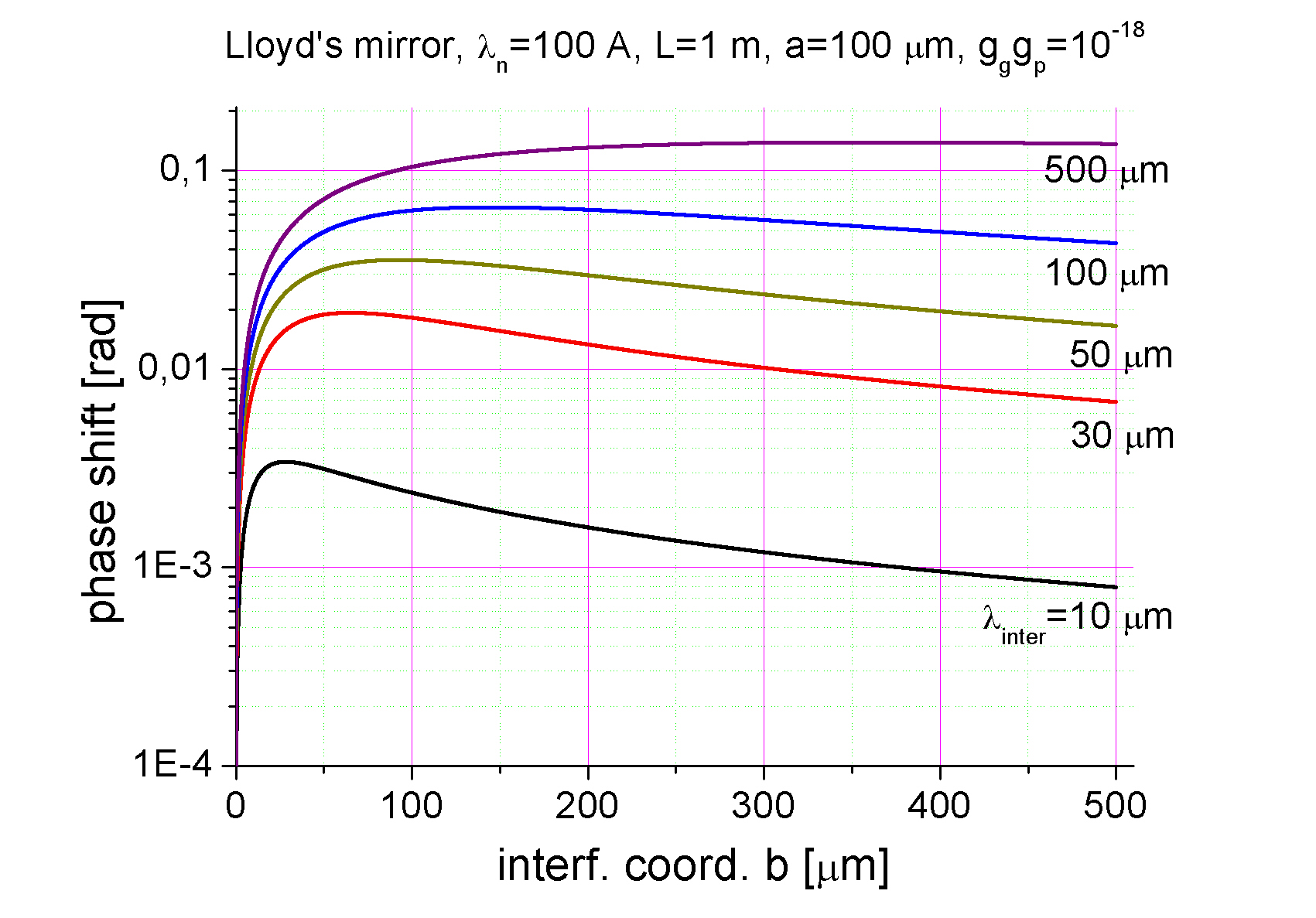}}
\end{center}
\caption{The neutron wave phase shift $\delta_{ax}$ at different interaction
range $\lambda_{inter}$ of the axion-like spin-dependent interaction with the
product of the coupling constants $g_{s}g_{p}=10^{-18}$.
 Lloyd's mirror interferometer has parameters the same as in Fig. 2.}
\end{figure}

\newpage

\begin{figure}
\begin{center}
\resizebox{13cm}{13cm}{\includegraphics[width=\columnwidth]{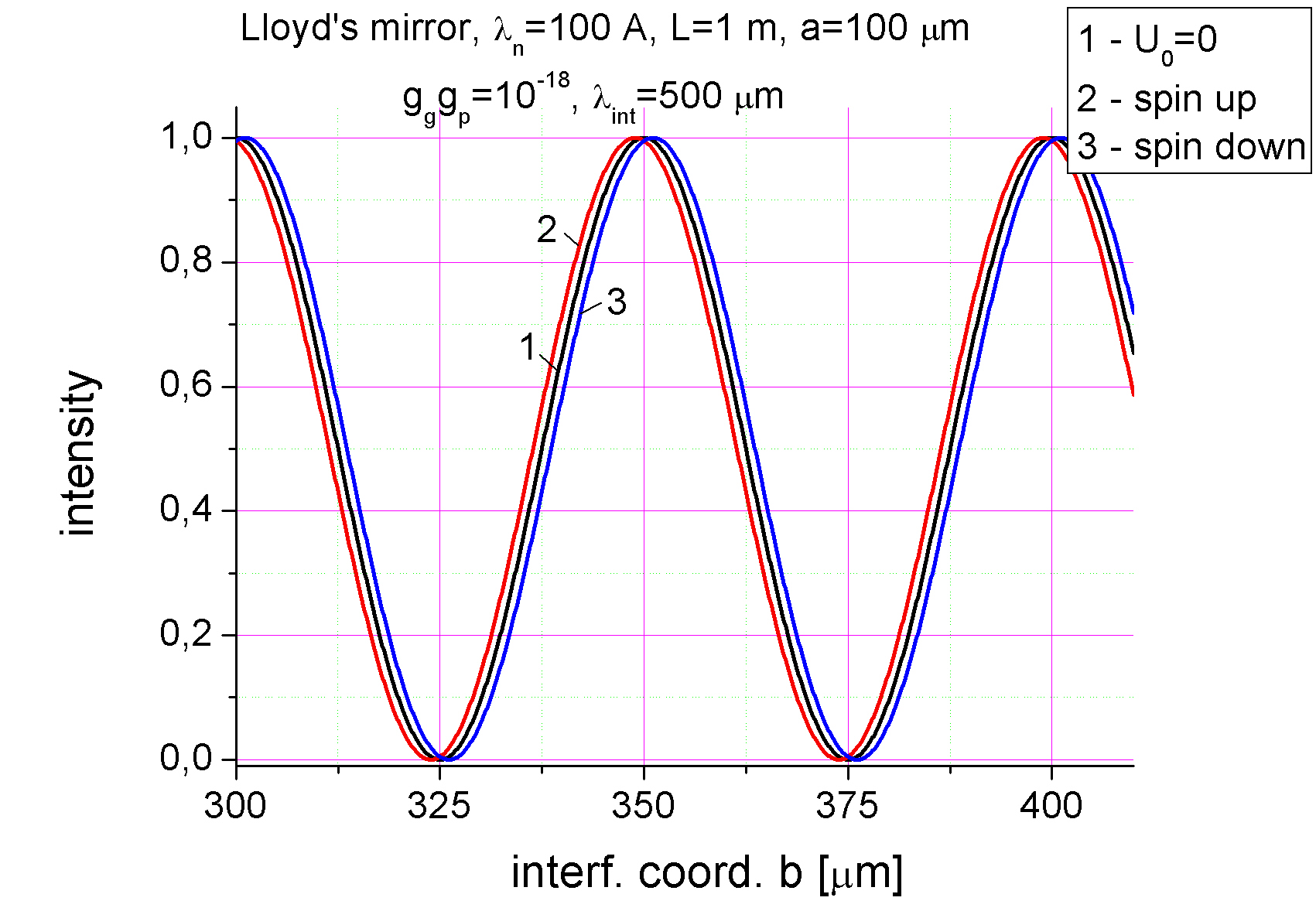}}
\end{center}
\caption{Calculated interference pattern due to the axion-like spin-dependent
interaction with $g_{s}g_{p}=10^{-18}$ and interaction range $\lambda=500 \mu$m.
 The interferometer has the same parameters as in Fig. 2.}
\end{figure}

\newpage

\begin{figure}
\begin{center}
\resizebox{13cm}{13cm}{\includegraphics[width=\columnwidth]{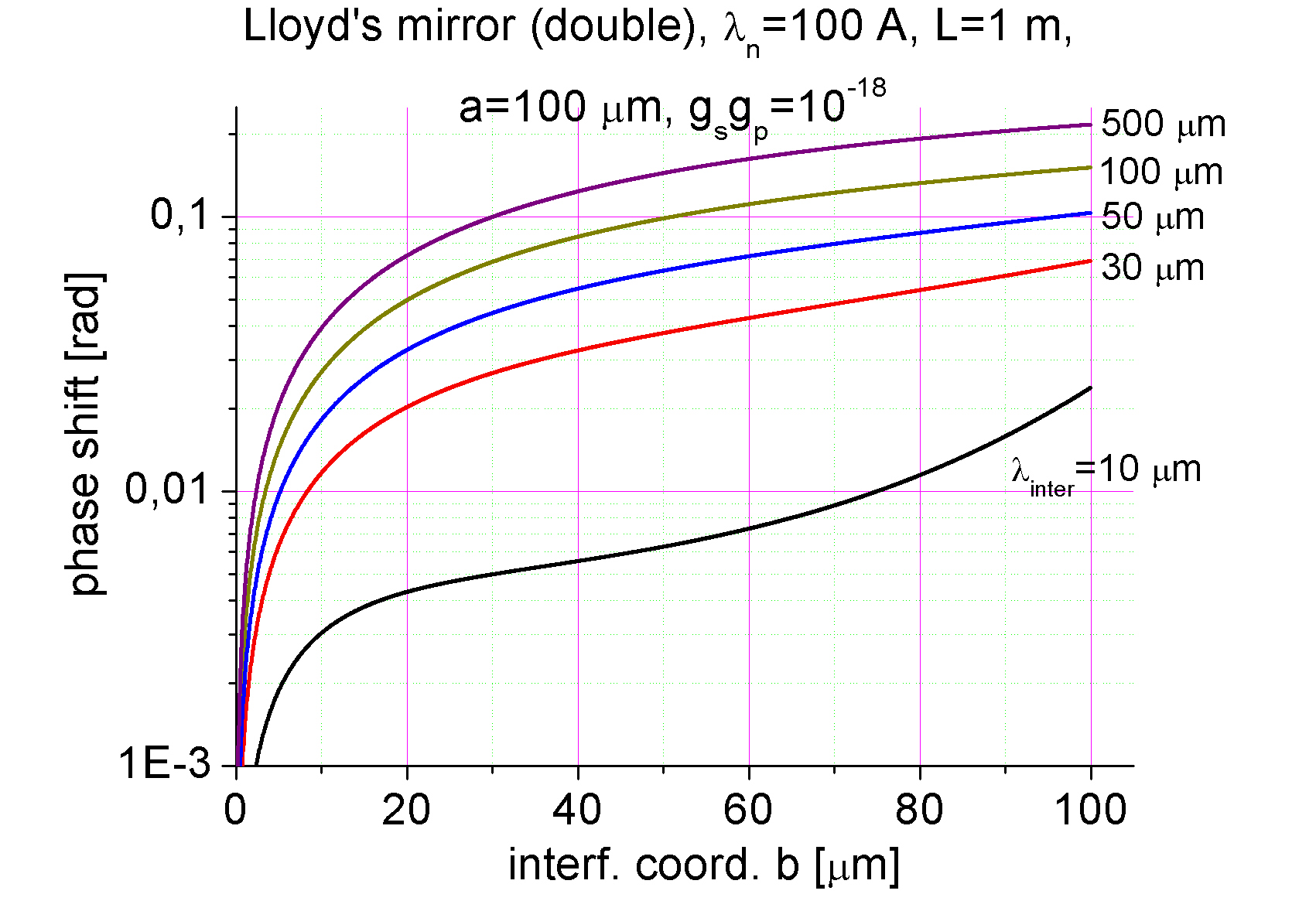}}
\end{center}
\caption{n The neutron wave phase shift $\varphi_{ax}$ at different interaction
range $\lambda_{inter}$ of the axion-like spin-dependent interaction with the
product of the coupling constants $g_{s}g_{p}=10^{-18}$.
 Lloyd's mirror interferometer has the geometry of Fig. 1(2) with the same
parameters as in Fig. 2, and the distance between mirrors 100 $\mu$m.}
\end{figure}

\newpage

\begin{figure}
\begin{center}
\resizebox{13cm}{13cm}{\includegraphics[width=\columnwidth]{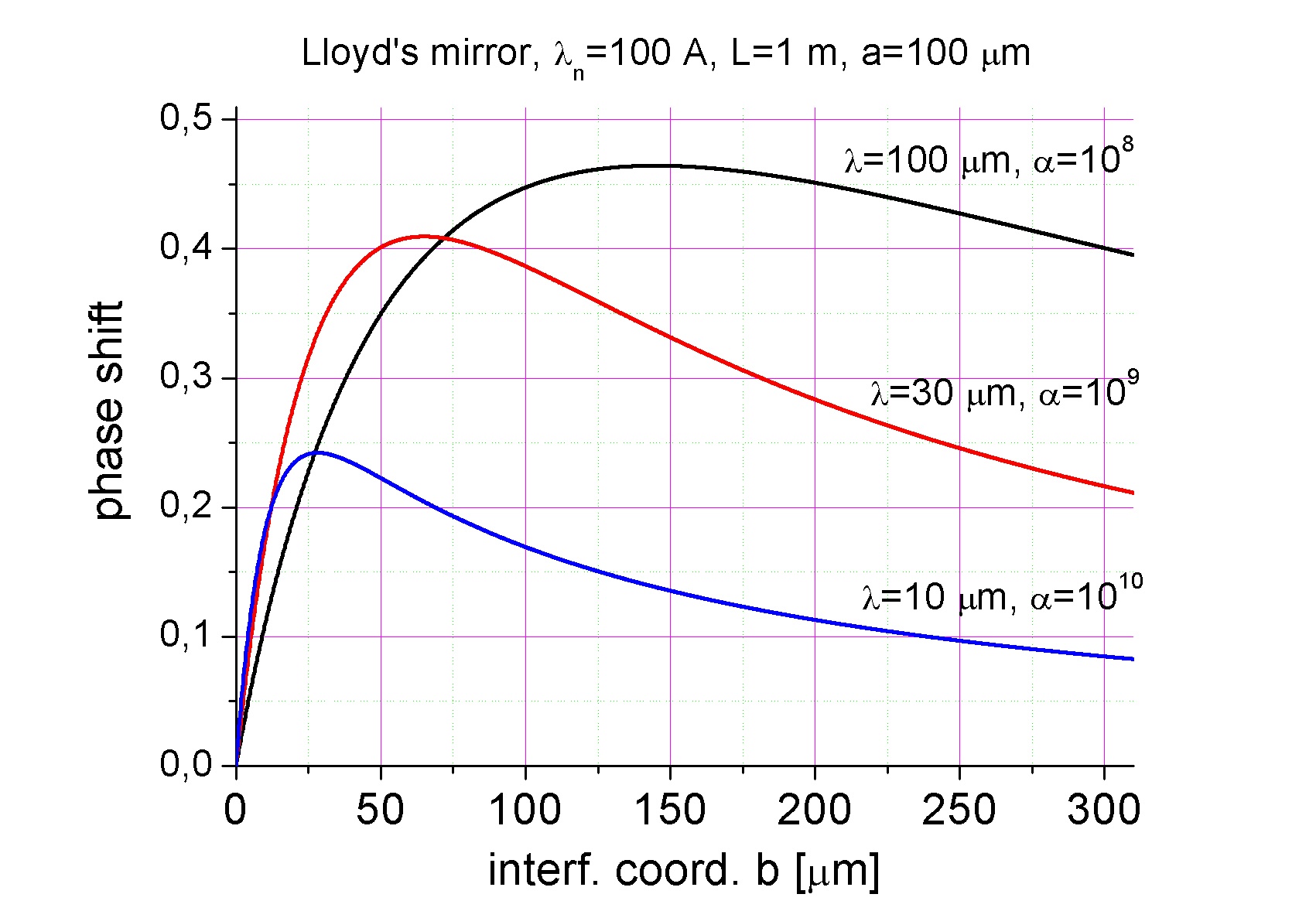}}
\end{center}
\caption{The neutron wave phase shifts $\varphi_{Yuk}$ in the Lloyd's mirror
interferometer of  geometry of Fig. 1(1) and with the same parameters as in
Fig. 2.}
\end{figure}

\begin{figure}
\begin{center}
\resizebox{13cm}{13cm}{\includegraphics[width=\columnwidth]{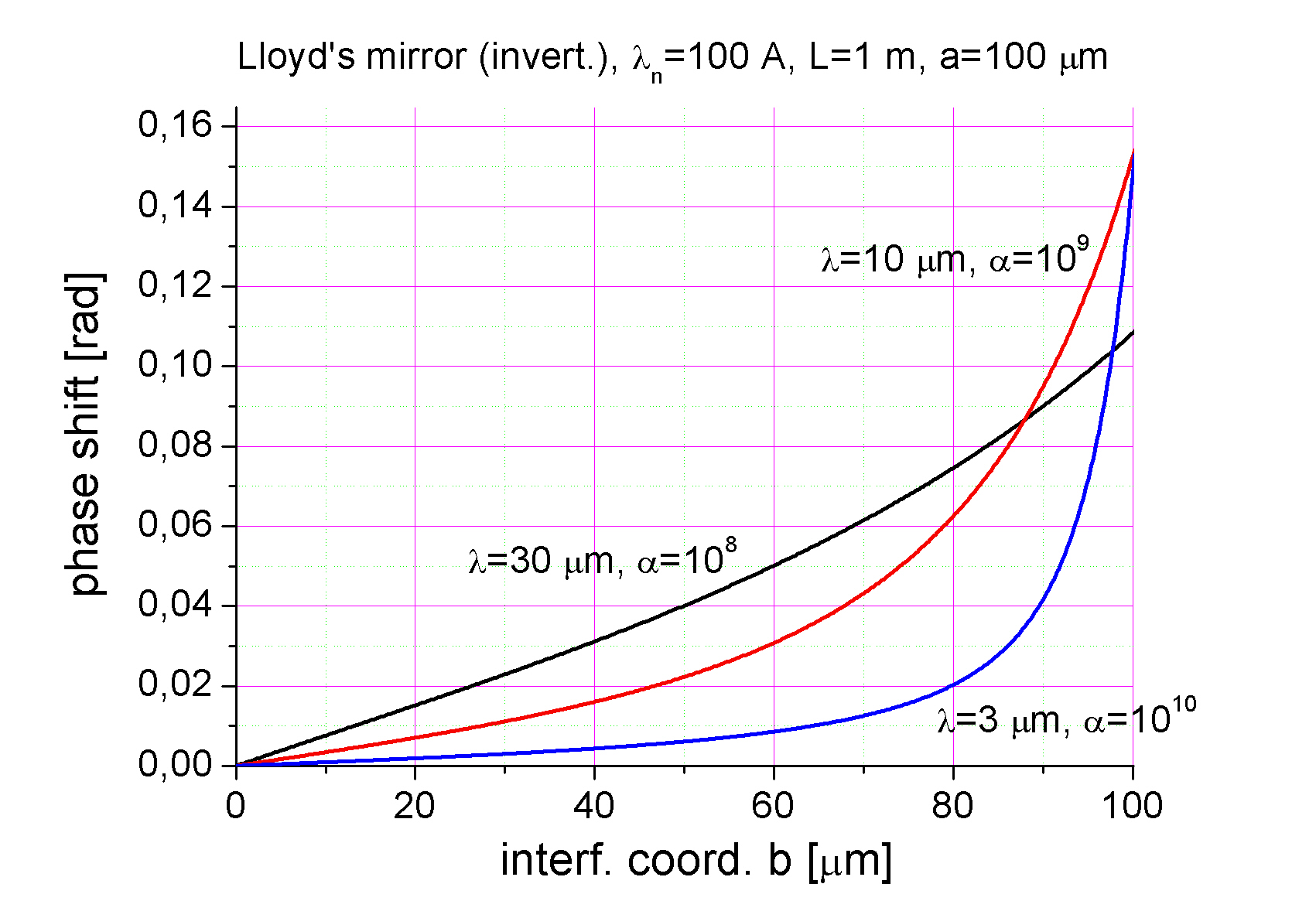}}
\end{center}
\caption{The neutron wave phase shifts $\varphi_{Yuk}$ in the Lloyd's mirror
interferometer of the geometry of Fig. 1(3) and with the same parameters as in
Fig. 2.}
\end{figure}

\end{document}